\documentstyle[11pt,newpasp,twoside,epsf]{article}
\begin{document}
\title{A New View on Young Pulsars in Supernova Remnants: Slow, Radio-quiet \& X-ray Bright}
 \author{E. V. Gotthelf}
\affil{Columbia University, 550 West 120$^{th}$ Street, New York, NY 10027, USA}
%Columbia Astrophysics Laboratory, 
\author{G. Vasisht}
\affil{Caltech/JPL, 4800 Oak Grove Drive, Pasadena, CA, 91109, USA}
%Jet Propulsion Laboratory, 

\begin{abstract}
We propose a simple explanation for the apparent dearth of radio
pulsars associated with young supernova remnants (SNRs).  Recent X-ray
observations of young remnants have revealed slowly rotating ($P \sim
10\rm{s}$) central pulsars with pulsed emission above 2 keV, lacking
in detectable radio emission.  Some of these objects apparently have
enormous magnetic fields, evolving in a manner distinct from the Crab
pulsar. We argue that these X-ray pulsars can account for a
substantial fraction of the long sought after neutron stars in SNRs
and that Crab-like pulsars are perhaps the rarer, but more highly
visible example of these stellar embers.  Magnetic field decay likely
accounts for their high X-ray luminosity, which cannot be explained as
rotational energy loss, as for the Crab-like pulsars.  We suggest that
the natal magnetic field strength of these objects control their
subsequent evolution.  There are currently almost a dozen slow X-ray
pulsars associated with young SNRs.
%these include the four known soft $\gamma$-ray repeaters, which have
%recently been confirmed as slow rotators. 
Remarkably, these objects, taken together, represent at least half of
the confirmed pulsars in supernova remnants. This being the case,
these pulsars must be the progenitors of a vast population of
previously unrecognized neutron stars.
\end{abstract}

\section{Where Are All The Young Neutron Stars?}

For the last 30 years it has been understood that young neutron stars
(NSs) are created during Type II/Ib supernova explosions involving a
massive star. Common wisdom holds that these neutron stars are born as
rapidly rotating ($\sim 10$ ms) Crab-like pulsars. Furthermore,
pulsars and their accompanying supernova remnants are thought to be
highly visible for tens of thousands of years, the former via
radio-loud, Crab-like ``plerionic'' (Weiler \& Sramek 1988) pulsar
nebulae, the latter as distinctive X-ray and radio shell-type
remnants. So where are all the young ($< 10^{4}$ yr) neutron stars? Of
the 220 known Galactic SNR (Green 1998) and over 1100 detected radio
pulsars (Camilo et~al., this volume), few associations between
the two populations have been identified with any certainty.

The current paradigm rests on the discoveries in the 1960's of the
Crab and Vela pulsars in their respective supernova 
nebulae. These were taken as spectacular confirmation for the
existence of neutron stars postulated much earlier by Baade \& Zwicky
(1934) based on theoretical arguments. The connection seems firm as
the properties and energetics of these pulsars could be uniquely
explained in the context of rapidly rotating, magnetized neutron stars
emitting beamed non-thermal radiation. Their fast rotation rates
and large magnetic fields ($\sim 10^{12}$ G) are consistent with those
of a main-sequence star collapsed to NS dimension and density. A fast
period essentially precluded all but a NS hypothesis and thus provided
direct evidence for the reality of NSs.
%(see Shapiro \& Teukolsky 1983 for a brief history and intro to NS physics). 
Furthermore, their
inferred age and association with SNRs provided strong evidence that
NSs are indeed born in supernova explosions.

So it is quite remarkable that, despite detailed radio searches, few
Galactic SNR have yielded a NS candidate over the years since the
initial discoveries. A recent census tallied only $10$ SNRs with pulsed
central radio sources (Helfand 1998). Furthermore, comprehensive radio
surveys suggest that most radio pulsars near SNRs shells can be
attributed to chance overlap (e.g. Gaensler \& Johnston 1995).
%; Kaspi et~al. 1996; Lorimer et~al. 1998). 
With the results of these new surveys, traditional
arguments for the lack of observed radio pulsars associated with SNR,
such as those invoking beaming and large ``kick'' velocities, become
less compelling, and perhaps even circular.

It is now clear that this discrepancy is an important and vexing
problem in current astrophysics. 
%Here we discuss an alternative
%evolutionary path for young NS which might account for the apparent
%lack of Crab-like pulsars associated with SNRs.

\section{The Revolution Evolution: Slowly Rotating Young X-ray Pulsars}

Progress in resolving this mystery is suggested by X-ray observations
of young SNRs. These are revealing X-ray bright, but radio-quiet
compact objects at their centers. It is now understood that these
objects form a distinct class of radio-quiet neutron stars (Caraveo et
al. 1996, Gotthelf, Petre, \& Hwang 1997 and refs. therein).
Often these objects have been labeled ``cooling neutron stars'', mainly
because of their lack of optical counterparts.

Some of these sources have been found to be slowly rotating pulsars
with unique properties. Their temporal signal is characterized by spin
periods in the range of $5 - 12$ s, steady spin-down rates, and highly
modulated sinusoidal pulse profiles ($\sim 30 \%$). They have steep
X-ray spectra (photon index $\ga 3$) with X-ray luminosities of
$\sim 10^{35}$ erg cm$^{-2}$~s$^{-1}$. As a class, these 
%seemingly isolated 
pulsars are currently referred to as the anomalous X-ray pulsars (AXP;
see refs. in Gotthelf \& Vasisht 1997).
% Duncan \& Thomson 1995; Mereghetti \& Stella 1995; van Paradijs, Taam \& van den Heuvel 1995). 
Nearly half are located at the centers of
SNRs, suggesting that they are relatively young ($\la 10^{5}$
yrs-old). And so far, no counterparts at other wavelengths have been
identified for these X-ray bright objects.  The prototype for this
class, the 7 s pulsar 1E 2259+586 in the $\sim 10^4$ yrs-old SNR
CTB~109, has been known for nearly two decades (Gregory \& Fahlman
1980). 

There are now almost a dozen slow radio-quiet X-ray pulsars
apparently associated with young SNRs. These include the four known
soft $\gamma$-ray repeaters (SGR) which have recently been confirmed
as slow rotators (Kouveliotou et~al. 1998), and likely associated with
young SNRs (e.g. Cline et al. 1982; Kouveliotou et~al. 1998).
%; Kulkarni \& Frail 1993; Vasisht et al. 1994), which have .
The census of these radio-quiet objects now approach in number those
estimated for those candidate young radio-bright objects connected with
SNRs.

\section{A Key Object: the Radio-quiet Slow X-ray Pulsar in Kes~73}

Given the latest X-ray results, it now appears likely that at least
half of the observed young neutron stars follow an evolutionary path
quite distinct from that of the Crab pulsar. An understanding of such
alternative paths for young NS evolution is suggested by
1E~1841$-$045, the remarkable 12-s anomalous X-ray pulsar in the
center of the SNR Kes 73. This young object has the longest period and
most rapid spin-down rate of any known isolated young pulsar. A recent
comprehensive study of the long term spin history of 1E~1841$-$045
indicated steady braking on a timescale of $\tau_s \simeq 4\times
10^3$ yrs, consistent with the inferred age of Kes~73 (Vasisht
et~al. 2000). The similarity in age along with the central location of
the pulsar strongly suggests that the two objects are related.

If the Kes 73 pulsar and other NS candidates like them were indeed
born as fast rotators, then a mechanism must be found to slow them
down to their currently observed rates. The rapid but steady spin-down
of the Kes~73 pulsar suggests a possibility. The equivalent magnetic
field for a rotating dipole is $B_{dipole} \simeq
3.2\times10^{19}~(P\dot P)^{1/2} \approx 8\times 10^{14}$ G, one of
the highest magnetic fields observed in nature.  Theory describing a
NS with such an enormous field, a ``magnetar'', has been worked out by
Duncan \& Thompson (1992). Vasisht \& Gotthelf (1997) suggest that the
Kes~73 pulsar was born as a magnetar $\sim 2\times10^3$ years ago and
has since spun down to a long period due to rapid dipole radiation
losses. The pulsar in Kes~73 provided the first direct evidence for a
magnetar through spin-down measurements and the apparent consistency in
age between the pulsar and SNR (see Gotthelf, Vasisht \& Dotani 1999
for details).

\begin{figure}
\epsfxsize=6.0cm
\centerline{\hfill \epsfbox[50 50 500 500]{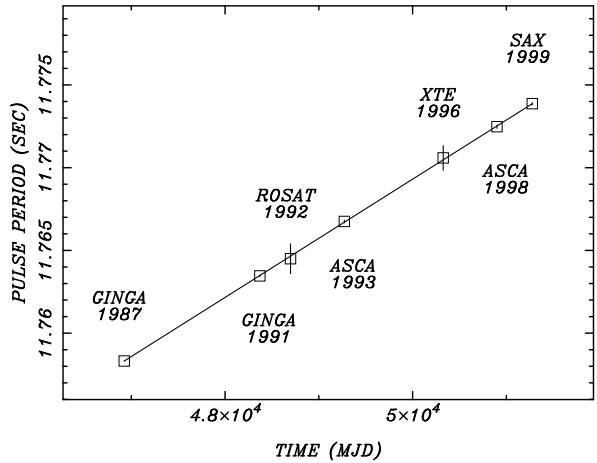}  \hfill}
\caption{Spin history of the 12-s central pulsar in the Galactic SNR
Kes~73. The rapid but steady spin-down, spanning over a decade,
provides the strongest evidence yet for a magnetar.  }
\end{figure}

In the magnetar model, the enormous magnetic field provides a natural
mechanism for braking the pulsar and spinning it down so
quickly. Pulsar spin-down via mass accretion, for any reasonable
accretion rate, would require longer than a Hubble time to spin-down
to the observed values. Furthermore, the observed luminosities of
these slow X-ray pulsars is consistent with them being powered by
magnetic field decay. If the total X-ray emission were powered by
rotational energy loss, as it the case for the radio pulsars, the
available energy is far too small. The maximum luminosity derivable
just from spin-down is $L_X \la 4\pi^2I \dot P/ P^3 \sim 10^{34} \ {\rm
erg \ s^{-1}}$ well below the measured value of $L_X \sim
3\times10^{35} \ {\rm erg \ s^{-1}}$.  On the other hand, the measured
luminosity is appreciably low for an accretion powered binary system
$\sim 10^{36-38} \ {\rm erg \ s^{-1}}$. These facts along with a lack
of stochastic variability and a steep spectrum makes an accretion
scenario all but unlikely.

In conclusion, it now seems likely that at least half the population
of young neutron stars in SNR evolve as slow AXP-like pulsars, as
exemplified by Kes~73. The Crab-like pulsars, highly visible via their
radio nebulae, are thus a less common manifestation of young NS
evolution.  We note that we do not need to invoke a substantial space
velocity for the former NSs, as those X-ray pulsars within known SNRs
typically lie at their centers.  The SGRs may represent an
evolutionary stage during which young NSs are likely to be produce
bursts. Under this scenario, the AXPs and SGR phenomena are closely
related, linked by their strong magnetic field.  We consider that many
of the young NSs ``missing'' in radio surveys can be accounted for by
the above discussed radio-quiet NSs.  As their evolution along the $P
- \dot P$ diagram cannot intersect the bulk of the aged radio pulsar
phase-space, AXP-like pulsars thus require the existence of a vast
population of previously unappreciated NSs.

\acknowledgments This research is supported by the NASA LTSA grant
NAG 5-7935.  E.V.G. thanks Steven Lawrence for comments.


\begin{references}
\reference{} Baade, W. \& Zwicky, F. 1934, Phys. Rev., 45, 138
\reference{} Caraveo, P. A., Bignami, G. F., Trumper, J. 1996, AARv, 7, 209
\reference{} Cline, T. L. 1982, ApJ, L255, 45
\reference{} Duncan. R. C. \& Thompson, C. 1992, ApJ, 392, 9
\reference{} Gaensler, B. \& Johnston, S.  1995, MNRAS, 277, 1243
\reference{} Gotthelf, E. V. \& Vasisht, G. 1997,  ApJ, 486, L133
\reference{} Gotthelf, E. V., Petre, R. \& Hwang, U. 1997, ApJ, 487, L175
\reference{} Gotthelf, E. V., G. Vasisht, \& Dotani, T. 1999, ApJ, 522, L49
\reference{} Gregory, P. C. \& Fahlman, G. G. 1980, Nature, 287, 805 
\reference{} Green D.A., 1998, http://www.mrao.cam.ac.uk/surveys/snrs
\reference{} Helfand, D. J. 1998, Mem. Soc. Astron. Ital., 69, 791
\reference{} Kouveliotou, C. et~al. 1998, Nature, 391, 235
\reference{} Vasisht, G. \& Gotthelf, E. V. 1997, ApJ, 486, L129 
\reference{} Vasisht, G., Gotthelf, E. V., Torii, K., \& Gaensler, B. 2000, in press
\reference{} Weiler, K. W. \& Sramek, R. A. 1988, ARA\&A, 26, 29.
\end{references}
\end{document}